\documentclass[9p,twocolumn]{revtex4-1}

\usepackage{leftidx}
\usepackage{amsmath}
\usepackage{tensor}
\usepackage{graphicx,epstopdf}

\usepackage{dcolumn}
\usepackage{bm}
\usepackage{hyperref}
\usepackage{amssymb}
\usepackage{hyperref}
\RequirePackage[normalem]{ulem} 
\RequirePackage{color}\definecolor{RED}{rgb}{1,0,0}\definecolor{BLUE}{rgb}{0,0,1} 

\begin{document}
\newcommand{\be}{\begin{equation}}
\newcommand{\ee}{\end{equation}}
\newcommand{\ba}{\begin{eqnarray}}
\newcommand{\ea}{\end{eqnarray}}
\newcommand{\Gam}{\Gamma[\varphi]}
\newcommand{\Gamm}{\Gamma[\varphi,\Theta]}
\thispagestyle{empty}

\title{Entanglement in helium atom confined in an impenetrable cavity}

\author{Przemys\l aw Ko\'scik$^{1}$}
\author{Jayanta K. Saha$^{2}$}
\address{$^{1}$Institute of Physics,  Jan Kochanowski University,
ul. \'Swi\c{e}tokrzyska 15, 25-406 Kielce, Poland}
\address{$^{2}$Indian Association for the Cultivation of Science,
Jadavpur, Kolkata 700032, India}

\begin{abstract}
We explore ground-state entanglement properties of  helium atom
confined at the center of  an impenetrable 
spherical cavity of varying
radius 
 by using explicitly correlated Hylleraas-type basis set.
Results for the dependencies   of the   von Neumann and
linear entanglement entropic measures on 
the cavity radius are discussed in details. Some highly accurate  numerical results  for
the von Neumann and linear entropy  are  reported for the first
time. It is found that the transition to the strong confinement
regime is manifested by the entropies as an appearance of the
inflection points on their 
variations.
\end{abstract}

\maketitle

\section{Introduction}
Study on the entanglement between identical interacting quantum particles is one of the rapidly  topic of the present day research. The knowledge or the understandings that comes out from such investigations directly enriches the underlying principles of the quantum information science \cite{niel} and related disciplines. On the other hand, due to the philosophical mapping between entanglement and correlations between the quantum particles, it has becoming an essential part of the structure
calculations of various microscopic
systems. As a consequence, last few years produced bulk of studies in this direction. Generally, the von Neumann (vN) entropy \cite{ghir} and the linear entropy \cite{manf}  are  used to
quantify the amount of entanglement in
composite quantum systems. In the most recent years, many  attempts have
been made towards understanding entanglement properties of real
two-electron systems, i.e., the helium atoms and helium-like ions
 \cite{mm,ho1,ho2,ho3,majo,desa1,desa2,bene1,kos1,res,hi,helj,helj1}.
 Related studies of entanglement in the context of  two-electron model systems such as  systems  of
 electrons confined by a spherical harmonic potential without and with an on-centre Coulomb impurity and the helium atom immersed
 in weakly coupled Debye plasmas are also available in the literature, \cite{koe}, \cite{kos2} and \cite{ho4} 
 respectively. In a recent study \cite{sah}, performed by both of us, the ground state entanglement properties of Helium atom  confined in a two parameter finite spherical cavity
 were also investigated.
 For an overview of recent progress in entanglement
studies of quantum composite systems, see \cite{tich}.  \\

The effective interaction between two electrons are modified while 
an atom is placed under different external environments e.g. plasma,
quantum dot (QD), fullerene cage etc. As a consequence, structural
and spectral atomic properties are being changed depending on the
type of the confining model potential representing the environment.
 Various phenomenological potentials exist in the literature \cite{akhi,ichi,xu,jia} which can mimic the corresponding environment.
 Historically the idea of confined atomic systems was first introduced by Michels et al. \cite{mich}, soon after the advent of quantum mechanics.
  Subsequently, different groups around the globe encountered this problem from different perspectives.
 Moreover sophisticated laboratory experiments on such systems in the last decade accelerated theoretical interest in recent times.
 A bulk of these studies, both theoretical and experimental, are well reviewed in some of the recent articles \cite{can,sab}.\\

In thermodynamic equilibrium, atoms under pressure confinement can
efficiently be modeled by an impenetrable spherical cavity under
adiabatic condition \cite{bhat}. Different properties of matter
under high pressure in Zovian planets \cite{guil} or in case of
geological sieves \cite{jaco} can be explained by using such
concepts of impenetrable confinement. Further modification of the
inter-particle interaction in this model can successfully be used
for atoms under dense plasma \cite{sbha}, endohedral atoms \cite{xu} etc.\\

In the present work, the model on which we focus is a helium atom
confined by impenetrable spherical cavity of radius $R$. Although a
lot work has been done in recent years in order to investigate
quantum properties of such system
\cite{bhat,con1,con2,con3,con4,con5,con8,con6}, its  entanglement properties  have not been  studied in the literature so far. 
We undertake this issue  and estimate  
both the vN and linear
entropy as a function of $R$. We have used the Hylleraas-type basis  to
incorporate the electron correlation in an explicit manner. Radius of
the cavity $R$ is varied from a high value mimicking the free system
to a low one corresponding to strong confinement regime.
The details of the present methodology is given in section 2
followed by discussion on the results obtained in section 3 and
finally concluded in section 4.

\section{Method}\label{12}
The Hamiltonian [atomic unit is used throughout] of a helium atom
confined in an impenetrable spherical cavity of radius $R$ is given
by \be
H=\sum_{i=1}^{2}\left[-\frac{1}{2}\nabla_{i}^{2}-\frac{2}{r_{i}}+V_{R}(r_{i})\right]+\frac{1}{r_{12}}
\label{ham},\ee where $V_{R}(r)= 0$ for $0 \leq r \leq R$ and
$\infty$ for $r> R$. In this  Letter we restrict our investigation
to the singlet ground-state of (\ref{ham}). Being the S-symmetry
state, its spatial wavefunction is expressed in terms of     $r_{1},
r_{2}$ and the  inter-electronic angle coordinate $\theta$,
$\psi(\textbf{r}_{1},\textbf{r}_{2})\equiv\psi(r_{1},r_{2},\cos\theta)$,
and   vanishes for $r_{1},r_{2}\geq R$. The Schmidt
decomposition of such a wavefunction takes  a form
 \cite{1,2}
\begin{eqnarray}\psi(\textbf{r}_{1},\textbf{r}_{2})=\sum_{{n,l=0}}^{\infty}\sum_{m=-l}^{m=l}{4\pi k_{nl}\over
2l+1} u_{nlm}^{*} (\textbf{r}_{1})u_{nlm}
(\textbf{r}_{2}),\label{Schmid}
\end{eqnarray}
 with \be u_{nlm}(\textbf{r})= {v_{nl}
(r)Y_{lm}(\theta,\varphi)\over r},\label{Shmid} \ee where $Y_{lm}$
are the spherical harmonics. Since the boundary condition imposed  on
the wavefunction: $\psi=0$, whenever $r_{i} = R, i = 1, 2$, the
radial orbitals $v_{nl} (r)$  meet the Dirichlet boundary condition:
 $v_{nl}(R)=0$ and together  with the coefficients $k_{nl}$
can be determined by the following integral equations \cite{kos2,1}
\be \int_{0}^{R} f_{l}(r,r^{'}) v_{nl} (r^{'}) dr^{'}=k_{nl}
v_{nl}(r),\label{integr}\ee with
 \begin{eqnarray}
f_{l}(r,r^{'})=\nonumber\\r r^{'}{2l+1\over 2}
\int_{0}^{\pi}\psi(r,r^{'},\cos\theta)P_{l}(\cos\theta)\sin\theta
d\theta=\nonumber\\=r r^{'}{2l+1\over 2}\int_{-1}^{1}\psi(r,r^{'},t)P_{l}(t)dt,\label{poloi1}
\end{eqnarray}
where  the upper limit of integration  is   $R$ in contrast  to $\infty$ for the free atom,
and $P_{l}$ is the lth order Legendre polynomial.

We will  measure the entanglement in  the ground-state of the system
under consideration  by the vN and linear entropy,
 which in the case of the singlet states
 have the forms \cite{kos2,vn3}
 \be S_{vN}=-\mbox{Tr}[\rho\log_{2}\rho],\label{polo}\ee  
and
 \be L=1-\mbox{Tr}[\rho^{2}],\label{polo1}\ee
respectively, where $\rho$ is the spatial reduced density matrix
$$\rho(\textbf{r},\textbf{r}^{'})=\int
[\psi(\textbf{r},\textbf{r}_{1})]^{*}\psi(\textbf{r}^{'},\textbf{r}_{1})d\textbf{r}_{1},$$
 the eigevectors  of which are  nothing but the Schmidt orbitals
(\ref{Shmid}) with eigenvalues $\lambda_{nl}$  related to the coefficients $k_{nl}$
 via the formula $\lambda_{nl}=({4\pi k_{nl}\over 2l+1})^2$ \cite{kos2,2}.
As the $2l+1$- fold degeneracy occurs, i.e., the eigenvectors
$\{u_{nlm}(\textbf{r})\}_{m=-l}^{m=l}$ correspond to the same
eigenvalue $\lambda_{nl}$, the conservation of probability gives
$\sum_{n,l=0} (2l+1) \lambda_{nl}=1$. As a result, in terms of
$\lambda_{nl}$, the vN and linear entropy  take the forms
 $S_{vN}=-\sum_{n,l=0} (2l+1) \lambda_{nl} \log_{2} \lambda_{nl}$ and
$L=1-\sum_{n,l=0} (2l+1) \lambda_{nl} ^2$, respectively.

\section{Results and Discussions}\label{12}

It is well known that the Ritz variational method in the framework
of Hyllerass-type basis and its variants are the best techniques
for precise numerical estimation of non-relativistic energy eigenvalues and
wavefunctions of
 two-electron systems. In our research for finding  the ground-state  energies and
wavefunctions of (\ref{ham}), we  apply an ansatz of Hyllerass-type
in the form \cite{hyl}:

\begin{eqnarray} \psi^{Hyl}=\left[R-{1\over 2}(s+t)\right]\left[R-{1\over 2}(s-t)\right]\nonumber\\\times\sum_{nmp} c_{nmp}e^{-\alpha s
}s^{n}t^{2m}u^{p}, \label{ff}\end{eqnarray}
 with  $0\leq n+m+p\leq \omega$, where $s, t$ and $u$ are the Hylleraas coordinates: $s=r_{1}+r_{2}, t=r_{1}-r_{2},
u=r_{12}=(r_{1}^2+r_{2}^2-2 r_{1} r_{2} \cos \theta)^{{1\over 2}}$, $\alpha$ is a non-linear variational parameter,
and the cut-off factors $R-{1\over 2}(s+ t)=R-r_{1}$; $R-{1\over
2}(s- t)=R-r_{2}$ are introduced to guarantee the correct
fulfillment of the boundary condition  at the hard walls. In each
case,
 the $\omega$-order approximations to the true  ground-state energy value, $E_{0}^{(\omega)}$, and the corresponding linear
parameters $C^{(\omega)}_{0}=[c_{ nmp}]$ are determined
  by solving the eigenproblem: \be
[H^{(\omega)}-E^{(\omega)}S^{(\omega)}]C^{(\omega)}=0, \ee where
$H^{(\omega)}=[\langle n^{'}m^{'}p^{'}|H|nmp\rangle]$,
 $S^{(\omega)}=[\langle n^{'}m^{'}p^{'}|nmp\rangle]$ are the Hamiltonian matrix and the overlap matrix, respectively, and   the non-linear parameter $\alpha$ is optimized  in order to minimize
$E_{0}^{(\omega)}$, $\partial_{\alpha} E_{0}^{(\omega)} =0$.

Here we determine  the coefficients
$k_{nl}$ by solving the Eq. (\ref{integr}) through a discretization
technique.  Namely, we turn  it into an algebraic
problem by discretizing the variables $r$ and $r^{'}$ with equal
subintervals of length $\Delta r$ such that   the resulting
diagonalization of the matrix $[\Delta r f_{l}( {i} \Delta r,{j}
\Delta r )]_{n_{m}\times n_{m}}$, $\Delta
r=R/(n_{m}-1)$, $i, j=0,1,...,n_{m}-1$  provides the $n_{m}$th order approximations to the
$n_{m}$ lowest coefficients $k_{nl}$.
 With the  coefficients $k_{nl}$  determined in that way,   approximations to  the vN and linear entropy, i.e., $S_{vN}=-\sum_{n,l=0}^{n_{m}-1,l_{m}} (2l+1) \lambda_{nl} \log_{2} \lambda_{nl}$,
$L=1-\sum_{n,l=0}^{n_{m}-1,l_{m}} (2l+1) \lambda_{nl} ^2$, ($\lambda_{nl}=({4\pi k_{nl}\over
2l+1})^2$), can be obtained with  a desired accuracy by increasing the  expansion length $\omega$ in (\ref{ff}) and the cut-offs $n_{m}$, $l_{m}$  until the results become stable to that accuracy.
It is worth mentioning at this point that  the approximations to the
  true values of $k_{nl}$ can alternatively be  obtained by diagonalizing the
matrix $[\langle u_{i}(r_{1})|
f_{l}(r_{1},r_{2})|u_{j}(r_{2})\rangle]_{n_{m}\times n_{m}}$, where
$u_{i}(r)=\sqrt{{2/ R}}\sin( {i \pi r/ R})$, $i, j=1,...,n_{m}$. For
more details on this point, see \cite{mm,ho3}, wherein the free helium atom was treated in that way  with   a set of orthonormal Laguerre basis functions.

\begin{figure}[h]
\begin{center}

\includegraphics[width=0.52\textwidth]{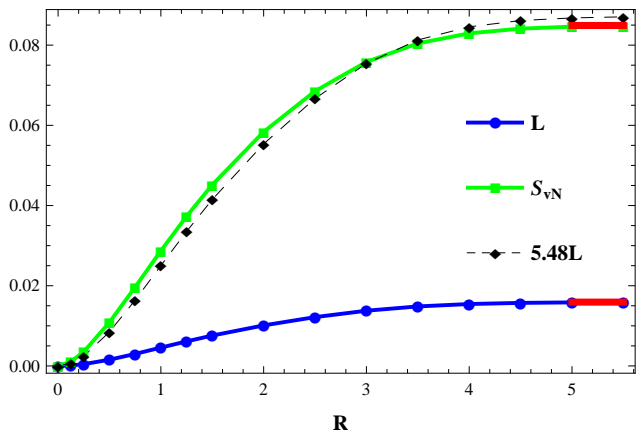}
\caption{ The ground-state vN and linear entropy together with the
rescaled linear entropy $5.48 L$ as a function of $R$. Horizontal
lines mark the results for the free helium atom
($R\rightarrow\infty$), $S_{vN}=0.08490, L=0.01592$ \cite{kos1}}
\label{fig:odog1}
\end{center}
\end{figure}

 For   demonstration of  convergence of the method, we present in table  \ref{tab:table11} the values obtained
 for  the ground-state energy and  the corresponding  linear
entropy at different  $R$
as a function of  $\omega$, where we also give the optimal values of $\alpha$.
      We found that the  results  for the  energy presented in this table   are in excellent agreement  with
      the already available data in the literature \cite{con4},   which confirms the correctness of our calculations.
                It is evident from the table
         that  the present variational scheme with  $\alpha$ optimized  so as to minimize $E_{0}^{(\omega)}$ is very efficient.
          Our numerical results accurate to the number of digits presented obtained for the ground-state at some representative values of $R$
          are summarized  in  table \ref{tab:table5}, where we also give the results  obtained in \cite{kos1} for the free helium atom.
         Fig. \ref{fig:odog1}  displays the dependencies of the ground-state  vN and linear entropy on $R$, where    a smaller cavity radius  at which  we have computed their  values is $R=0.125$.
We concluded from our calculations that a rate of convergence of the method with
respect to $l_{m}$  only slightly depends on $R$. 

 It turned out that, at least over the range of $R$ from $0.125$ to $\infty$,
the numerical stability of the results determined with $6$-digit
precision for the ground-state  vN and linear entropy   is achieved
 at  about $l_{m}=18$ and at  $l_{m}=1$, respectively.
For a detailed  convergence analysis  with respect to $l_{m}$
  in case of the ground-state  of free
 helium atom as $R\rightarrow\infty$ we   refer  the readers to \cite{kos1}.

 We also gained some insight into  the relation between the  vN  and  linear entropy
  as a function of $R$. Our analysis has revealed   that the  rescaled linear entropy  such that $5.48L$   behaves qualitatively in a similar
 way as the vN entropy, where the factor $5.48$  is obtained as the best
proportionality constant between our numerical data for the vN and
linear entropy. This is  demonstrated in  Fig. \ref{fig:odog1},
where the  variation of  $5.48L$ is also shown. Strictly speaking,
  the relative error of the  approximation $S_{vN}\approx5.48L$ is about $38$ percent at $R=0.125$ and   quickly decreases as $R$ is further  increased until it becomes    less than $2.7$ percent for  $R$ larger than $3$.

   It is apparent from our results that  already as the cavity radius $R$ exceeds   a value  of about  $R=5$, the ground-state of the  corresponding confined system  behaves almost as the ground-state of
    the free
    helium atom. As can be seen, the entanglement is the largest in the free
    atom regime  and decreases monotonically  as  $R$ decreases  until it vanishes in the limit of an infinitely strong
confinement, i.e., when  $R\rightarrow 0$. In this limit the electrons behave like noninteracting electrons
 and the corresponding state can be regarded as non-entangled.
 Interestingly enough,   Fig. \ref{fig:odog1} indicates    that the transition to the strong confinement regime is manifested
 by the behaviors of the vN and linear entropy as a change  of the signs of their second derivatives.
To gain a deeper   understanding of    this point, we constructed interpolation functions of our  numerical data  for the vN and linear entropy and
      computed the roots of their second derivatives,  ${d^2 L\over d R^2}=0$ and ${d^2 S_{vN}\over d R^2}=0$, which resulted in  the inflection points at  $R\approx0.81$ and at $R\approx0.95$, respectively.
      It is worthwhile to note that  these values are rather far away  from a value of  atomic ionization radius $R_{c}\approx1.385$ that is usually defined in the literature as that corresponding to the transition to the strong confinement regime \cite{con2,con6} (for $R$  below
$R_{c}$ the atom has a higher energy than $He^{+}$). However, the
closeness of the inflection points   to  a value of the critical
cavity radius at which the system gets completely  ionized, i.e.,
when  $E_{0} = 0$,  $R_{c} \approx 1.101$, is  worth noting.
Finally, there may be a general interest in noting    that in the
system under consideration  the entanglement entropies behave
qualitatively in the same way as the  Shannon information entropy in
position space. Namely, as it was found in  \cite{shan}, it also
monotonically increases   as the  radius of confinement increases
and saturates at a constant value as the cavity radius R exceeds a
value of about $R=5$.

\begin{table}[h]
\begin{center}
\begin{tabular}{llllll}
\hline
&  &$R=1$&$R=2.5$&$R=5$ \\

\hline
$\omega=3 $ &$\alpha$  & $1.8$&$1.1$&$1.5$ \\
 $ $ &$E_{0}^{(\omega)}$  & $1.0157602$&$-2.807762$&$-2.903333$ \\
 $ $ &$L^{(\omega)}$  & $0.00456808$&$0.0122032$&$0.0159295$\\

\hline

$\omega=4 $ &$\alpha$  & $2$&$1.1$&$1.5$ \\
 $ $ &$E_{0}^{(\omega)}$  & $1.0157556$&$-2.807820$&$-2.903383$ \\
 $ $ &$L^{(\omega)}$  & $0.00456831$&$0.0121797$&$0.0158641$ \\

\hline
$\omega=5 $ &$\alpha$  & $2$&$1.3$&$1.6$ \\
 $ $ &$E_{0}^{(\omega)}$  & $1.0157551$&$-2.807833$&$-2.903406$ \\
 $ $ &$L^{(\omega)}$  & $0.00456829$&$0.0121742$&$0.0158483$ \\

\hline
$\omega=6 $ &$\alpha$  & $$&$1.3$&$1.7$ \\
 $ $ &$E_{0}^{(\omega)}$  & $-$&$-2.807835$&$-2.903409$\\
 $ $ &$L^{(\omega)}$  & $-$&$0.0121738$&$0.0158449$ \\

\hline
$\omega=7 $ &$\alpha$  & $-$&&$1.8$\\
 $ $ &$E_{0}^{(\omega)}$  & $-$&${-}$&$-2.903411$ \\
 $ $ &$L^{(\omega)}$  & $-$&$-$&$0.0158434$ \\

\hline
\end{tabular}
\caption{\label{tab:table11}$\omega$-order approximations to the
ground-state energy ($E_{0}^{(\omega)}$) and the corresponding
linear entropy ($L^{(\omega)}$) determined  as
discussed in the text as a function of the expansion length $\omega$ for some values of
$R$}
\end{center}
\vspace{-0.6cm}
\end{table}

\begin{table}[h]
\begin{center}
\begin{tabular}{llllll}
\hline
$$ & $R=0.5$&$R=1$ &$R=2.5$&$R=5$& $R=\infty$ \\
\hline

$E_{0}$ &$22.7413 $& $1.01576 $&$-2.8078$&$-2.9034$& $-2.9037$ \\

$L$ &$0.00155 $& $0.00457 $&$0.01217$&$0.01584$& $0.01592$ \\

$S_{vN}$ &$0.01088$& $0.02871 $&$0.06852$&$0.08460$& $0.08490$ \\

\hline
\end{tabular}
\caption{\label{tab:table5} Ground-state energy and the
corresponding vN entropy ($S_{vN}$) and  the linear entropy ($L$)
obtained for some cavity radius $R$ are compared with the results
for the free helium atom }
\end{center}
\vspace{-0.6cm}
\end{table}

\section{Conclusions}\label{summ}
In conclusion, we have investigated the ground-state entanglement properties of helium atom  spherically enclosed by impenetrable boxes of varying   radius $R$. Both the vN and linear entropy have been computed in the  range of $R$ covering entirely their most rapid variations. In general, the presence
of the cavity  decreases  the entanglement for
every $R$, having its most pronounced   effect  in the range of values of $R$ smaller than  $R\approx5$. The entanglement entropies deviate more and more from those of the
free
 helium atom as the confinement becomes stronger and stronger. Our calculations
have shown  that the transition to the strong confinement
 regime  is manifested  by   inflections points appearing in the behaviors of the vN and linear entropy.
 We found that these points are  close to a critical cavity radius at which the system  gets completely ionized.


\begin{thebibliography}{99}

\bibitem{niel} M. A. Nielsen, I. L. Chuang, Quantum Computation and Quantum Information, Cambridge University Press, Cambridge, 2000.
\bibitem{ghir} G. Ghirardi, L. Marinatto, Phys. Rev. A \textbf{70}, 012109 (2004)
\bibitem{manf} G. Manfredi, M. R. Feix, Phys. Rev. E \textbf{62}, 4665 (2000)
\bibitem{mm} C. H. Lin, Y. K. Ho,Few Body Sys. \textbf{55}, 1141 (2014)
\bibitem{ho1} Y. C. Lin, C. Y. Lin, Y. K. Ho, Phys. Rev. A \textbf{87}, 022316 (2013)
\bibitem{ho2} C. H. Lin, C. Y. Lin, Y. K. Ho, Few Body Sys. \textbf{54}, 2147 (2013)
\bibitem{ho3} Y. C. Lin, Y. K. Ho, Phys. Lett. A \textbf{378}, 2861 (2014)
\bibitem{majo} D. Manzano, A. R. Plastino, J. S. Dehesa and T. Koga, J. Phys. A \textbf{43}, 275301 (2010)
\bibitem{desa1} J. S. Dehesa et al., J. Phys. B \textbf{45}, 015504 (2012)
\bibitem{desa2} J. S. Dehesa et al., J. Phys. B \textbf{45}, 239501 (2012)
\bibitem{bene1} G. Benenti, S. Siccardi, G. Strini, Eur. Phys. J. D \textbf{67}, 83 (2013)
\bibitem{kos1} P. Ko\'{s}cik, A. Okopi\'{n}ska, Few Body Sys. \textbf{55}, 1151 (2014)
\bibitem{res} J. P. Restrepo Cuartas and J. L. Sanz-Vicario, Phys. Rev. A \textbf{91}, 052301 (2015)
\bibitem{hi} C. H. Lin, Y. K. Ho, Few Body Sys. \textbf{56}, 157 (2015)
\bibitem{helj}S. L\'{o}pez-Rosa, et al.,  J. Phys. B: At. Mol. Opt. Phys. \textbf{48}  175002 (2015)
\bibitem{helj1} Y.C.Lin, Y.K. Ho, Can. J. Phys. \textbf{93}, 646-653 (2015)

\bibitem{koe} J. P. Coe, A. Sudbery, I. D'Amico, Phys. Rev. B \textbf{77}, 205122 (2008)
\bibitem{kos2} P. Ko\'{s}cik, Phys. Lett. A \textbf{377}, 2393 (2013)
\bibitem{ho4} Y. C. Lin, T. K. Fang, Y. K. Ho, Phys. Plas. \textbf{22}, 032113 (2015)
\bibitem{sah} P. Ko\'{s}cik, J. K. Saha, Few Body Sys. online published (2015)
\bibitem{tich} M. Tichy, F. Mintert, A. Buchleitner, J. Phys. B \textbf{44}, 192001 (2011)
\bibitem{akhi} A.I. Akhiezer et al., Plasma Electrodynamics, Vol. 1. Linear Response Theory (Oxford: Pergamon) 1975.
\bibitem{ichi} S. Ichimaru, Rev. Mod. Phys. 54, 1017 (1982)
\bibitem{xu} Y. B. Xu, M. Q. Tan, U. Becker, Phys. Rev. Lett. \textbf{76}, 3538 (1996)
\bibitem{jia} L. G. Jiao and Y. K. Ho, Electronic Structure of Quantum Confined Atoms and Molecules K.D. Sen (ed.) (2014) p145 .
\bibitem{mich} A. Michels, J. de Boer, A. Bijl, Physica \textbf{4}, 981 (1937)
\bibitem{can} S. Canuto (ed), Solvation Effects on Molecules and Biomolecules, Computational Methods and Applications (2008) (Berlin: Springer).
\bibitem{sab} J. Sabin, E. Brandas (ed), Adv. Quan. Chem. \textbf{57}, 1-334 (2009)
\bibitem{bhat} S. Bhattacharyya, J. K. Saha, P. K. Mukherjee, T. K. Mukherjee, Phys. Scr. \textbf{87}, 065305 (2013)
\bibitem{guil} T. Guillot, Planet Space Sci. \textbf{47}, 1183 (1999)
\bibitem{jaco} P. A. Jacobs, Carboniogenic Activity of Zeolites (Amsterdam: Elsevier) 1997.
\bibitem{sbha} S. Bhattacharyya, J. K. Saha, T. K. Mukherjee, Phys. Rev. A \textbf{91}, 042515 (2015)
\bibitem{con1} N. Aquino, A. Flores-Riveros, J. F. Rivas-Silva, Phys. Lett. A \textbf{307}, 326 (2003)
\bibitem{con2} H. E. Montgomery Jr., N. Aquino, A. Flores-Riveros, Phys. Lett. A \textbf{374}, 2044 (2010)
\bibitem{con3} A. Flores-Riveros, N. Aquino, H. E. Montgomery Jr.,  Phys. Lett. A \textbf{374}, 1246 (2010)
\bibitem{con4} A. Flores-Riveros, A. Rodriguez-Contreras, Phys. Lett. A \textbf{372}, 6175 (2008)
\bibitem{con5} C. L. Wilson, H. E. Montgomery Jr., K. D. Sen , D. C. Thompson,  Phys. Lett. A \textbf{374}, 4415 (2010)
\bibitem{con8} H. E. Montgomery Jr., Vladimir I. Pupyshev,Phys. Lett. A \textbf{377}, 2880 (2013)
\bibitem{con6} C. Laughlin and S. I. Chu, J. Phys. A: Math. Theor. \textbf{42}, 265004 (2009)
\bibitem{1} E. R. Davidson, The Journal of Chemical Physics, \textbf{39}, 875 (1964)
\bibitem{2} J. Wang, C. K. Law, and M.-C. Chu, Phys. Rev. A 72, 022346 (2005)
\bibitem{vn3} Schr\"{o}ter, S., H. Friedrich, and J. Madro\~{n}ero, Phys. Rev. A 87, 042507 (2013)
\bibitem{hyl} E. A. Hylleraas, Z. Phys. 54 (1929) 347, translated in H. Hettema, Quantum Chemistry World Scientific, Singapore, 2000.
\bibitem{shan}K. D. Sen, J. Phys. Chem. 123, 074110 (2005).
\end{thebibliography}
\end{document}